\documentclass[12pt]{article}
\usepackage{amsmath,amssymb,mathtools,cite,graphicx,fancyhdr,bbold,color}
\begin{document}


\title{The good, the bad and the ugly coherent states 
through polynomial Heisenberg algebras}
\lhead{The good, the bad and the ugly coherent states}
\rhead{Castillo-Celeita, Fern\'andez}

\author{Miguel Castillo-Celeita\footnote{{\it email:} mfcastillo@fis.cinvestav.mx} 
\ and David J. Fern\'andez C.\footnote{{\it email:} david@fis.cinvestav.mx}
\\ [10pt] {\sl Departamento de F\'isica, Cinvestav} \\ 
{\sl A.P. 14-740, 07000 Ciudad de M\'exico, M\'exico}}

\date{}

\maketitle

\begin{abstract}
Second degree polynomial Heisenberg algebras are realized through the harmonic oscillator 
Hamiltonian, together with two deformed ladder operators chosen as the third powers of the standard 
annihilation and creation operators. The corresponding solutions to the Painlev\'e IV equation are easily
found. Moreover, three different sets of eigenstates of the deformed annihilation operator are 
constructed, called the good, the bad and the ugly coherent states. Some physical properties of such 
states will be as well studied.
\end{abstract}

\section{Introduction}
\label{sec:1}
Polynomial Heisenberg algebras (PHA) of second degree are interesting deformations of the Heisenberg-Weyl 
algebra. In a differential representation they can be realized by one-dimensional Schr\"odinger 
Hamiltonians, together with a pair of third order ladder operators. In fact, when looking for the 
most general Hamiltonian ruled by such algebraic structure, it turns out that the potential depends on
solutions to a non-linear second-order ordinary differential equation called Painlev\'e IV (PIV) equation. 
Reciprocally, if one has Hamiltonians with third-order differential ladder operators, then it is 
possible to design a simple algorithm for generating solutions to such an equation, by identifying just 
the associated extremal states \cite{CFNN04,BF14}. 

On the other hand, it is important to look for the simplest systems ruled by second degree PHA, such that 
the corresponding extremal states satisfy the boundary conditions for being eigenfunctions of the Hamiltonian 
\cite{Ce15,CF16}. This is the main subject to be addressed in this work. Indeed, it will be shown that the 
harmonic oscillator Hamiltonian, together with deformed ladder operators which are the third powers of the 
standard annihilation and creation operators, will define a second degree PHA with such properties 
(Section~\ref{sec:2}). The three solutions of the PIV equation associated to this deformed algebra will be 
derived at the same Section. The corresponding coherent states (CS) as well as their properties, will be 
studied in Section~\ref{sec:3}, while Section~\ref{sec:4} will contain our conclusions.

\section{Second degree PHA for the harmonic oscillator}
\label{sec:2}

\begin{figure}
\begin{center}
\includegraphics[scale=1.1]{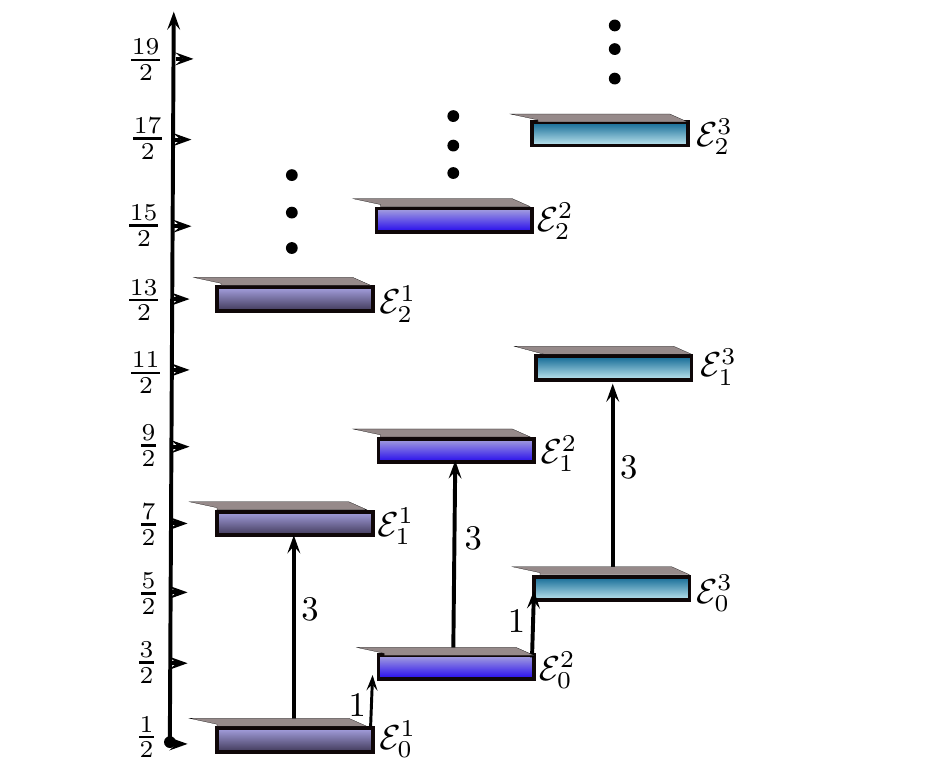}
\end{center}
\caption{The three independent ladders (with spacing $\Delta E=3$) 
for the second degree PHA of Eq.~(\ref{hoHam}).
They produce globally the harmonic oscillator spectrum with the standard spacing 
$\Delta E=1$.}
\label{twoladders}
\end{figure}

There are several ways to realize the second degree PHA through the harmonic oscillator. Here, we look for 
realizations such that the three extremal states are eigenfunctions of $H$ and, thus, we can generate from 
them three infinite ladders of eigenfunctions and eigenvalues \cite{Ce15}. Let us consider then the 
deformed ladder operators,
\begin{eqnarray}
& a_g = a^3, \qquad a_g^+ = (a^+)^3 .
\end{eqnarray}
The operator set $\{H, a_g, a_g^+\}$ gives place to a second degree PHA, since
\begin{eqnarray}
& [H,a_g] =  - 3 a_g, \label{hoHam} \qquad [H,a_g^+] = 3 a_g^+,  \label{hoHap} \qquad
[a_g,a_g^+] = N(H + 3) - N(H), \label{hoamap}
\end{eqnarray}
where the analogue of the number operator reads:
\begin{eqnarray}
& N(H) = a_g^+ \, a_g = \left(H - \frac12\right) \left(H - \frac32\right) \left(H - \frac52\right).
\end{eqnarray}
Three extremal state energies are identified, ${\cal E}_j = E_{j-1} = j-\frac12, \ j=1,2,3$,
with eigenvectors given by:
\begin{eqnarray}
&& \vert \psi_{{\cal E}_j} \rangle \equiv \vert \psi_{0}^j \rangle = \vert j-1 \rangle, \quad j=1,2,3,
\end{eqnarray}
where $\vert j-1 \rangle, \ j=1,2,3$ are the first three energy eigenstates of the harmonic oscillator 
in Fock notation. Departing from them, by acting $a_g^+$ iteratively, we can construct three independent 
ladders of energy eigenstates. The eigenvalues associated to the $j$-th ladder are
${\cal E}_{n}^j = {\cal E}_j + 3n, \ n=0,1,\dots, \ j=1,2,3,$
and the corresponding eigenstates become
\begin{eqnarray}
& \vert \psi_{n}^j \rangle = \vert 3n + j - 1 \rangle = 
\sqrt{\frac{(j-1)!}{(3n+j-1)!}}(a_g^+)^n \vert j - 1 \rangle, \quad j=1,2,3.
\end{eqnarray}
The spectrum of $H$ thus takes the form:
\begin{eqnarray}
& {\rm Sp}(H) = \{{\cal E}_{0}^1, {\cal E}_{1}^1,\dots \} \cup 
\{{\cal E}_{0}^2, {\cal E}_{1}^2,\dots \} \cup 
\{{\cal E}_{0}^3, {\cal E}_{1}^3,\dots \},
\end{eqnarray}
which is the harmonic oscillator spectrum seen from a new viewpoint: the Hilbert space 
is the direct sum of three orthogonal supplementary subspaces, ${\cal H} = {\cal H}_0 
\oplus{\cal H}_1 \oplus{\cal H}_2$, each one of them containing one ladder, which is represented in 
Figure~\ref{twoladders}. 

Since $\{H, a_g, a_g^+\}$ generate a second degree PHA, there is a link with the PIV equation 
\cite{CFNN04,BF14}:
\begin{eqnarray}
&& \frac{d^2g}{dy^2} = \frac{1}{2g}\left(\frac{dg}{dy}\right)^2 + \frac32 g^3 + 4yg^2+2(y^2-a)g 
+ \frac{b}{g},
\end{eqnarray}
which allows us to find some of its solutions. We just need to supply the three extremal
states and their associated energies, in our case $\psi_{{\cal E}_j}(x) = \langle x\vert
j-1\rangle, \ {\cal E}_j = j-1/2, \ j=1,2,3$. The PIV solution and its parameters 
turn out to be given by:
\begin{eqnarray}
&& \hskip-0.5cm g(y) = - y -\frac{d}{dy}\left[\ln \phi_1(y)\right], \quad a = \tilde{\cal E}_2 +
\tilde{\cal E}_3+2 \tilde{\cal E}_1-1, \quad b= -2\left(\tilde{\cal E}_2 - \tilde{\cal E}_3\right)^2,
\end{eqnarray}
where $\phi_1(y)$ is the first extremal state for the previous ordering (the ground state), and 
$y = \sqrt{3}x$, $\tilde{\cal E}_j = {\cal E}_j/3, \ j=1,2,3$ are the changes required to fit the 
spacing of levels of our system ($\Delta E=3$) with the standard spacing ($\Delta E=1$) used in 
\cite{CFNN04,BF14}. Since the first label can be asigned to any extremal state, we can find indeed three PIV solutions, 
whose explicit expressions and corresponding parameters become:
\begin{eqnarray}
& g(y) = -2y/3,  \qquad a=0, \qquad b=-2/9, \\
& g(y) = -2y/3 -1/y, \qquad a=-1, \qquad b=-8/9, \\
& g(y) = -2y/3 -4 y/(2y^2-3), \qquad a=-2, \qquad b=-2/9.
\end{eqnarray}

\section{Coherent states}
\label{sec:3}

Let us consider now the CS as eigenstates of the deformed annihilation operator:
\begin{eqnarray}
& a_g \vert \alpha \rangle_j = \alpha \vert \alpha \rangle_j, \quad j=0,1,2,
\end{eqnarray}
with
$\vert \alpha \rangle_j = \sum\limits_{n=0}^\infty C_n \vert 3n + j \rangle .$
Following a standard procedure, we arrive at:
\begin{eqnarray}
& \vert \alpha \rangle_j =
\frac{1}{\sqrt{\sum\limits_{n=0}^{\infty}\frac{|\alpha|^{2n}}{(3n+j)!}}}
\sum\limits_{n=0}^{\infty}\frac{\alpha^n}{\sqrt{(3n+j)!}}\vert 3n + j \rangle .
\end{eqnarray}

\begin{figure}
\begin{center}
\includegraphics[scale=0.75]{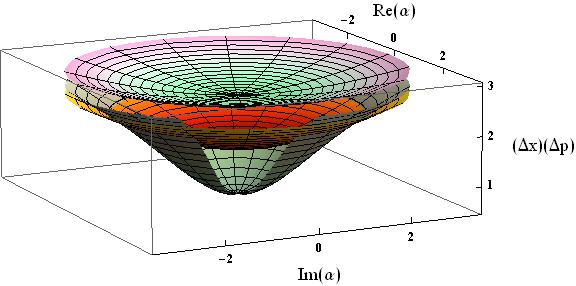}
\end{center}
\caption{Uncertainty products $(\Delta x)_j(\Delta p)_j$; 
the minima are $\frac12$, $\frac32$ and $\frac52$ for $j=0,1,2$, respectively.}
\label{dxdpj0}
\end{figure}

Several important quantities for the CS $\vert \alpha \rangle_j$ can be obtained 
straightforwardly:
\begin{eqnarray}
& \langle x\rangle_j =  \langle p\rangle_j = 0 , \quad 
\langle x^2\rangle_j = \langle p^2\rangle_j  = (\Delta x)_j(\Delta p)_j = \langle H \rangle_j =
\vert a \vert \alpha \rangle_j \vert^2 + \frac12,
\end{eqnarray}
where
\begin{eqnarray}
& \vert a \vert \alpha \rangle_j \vert^2 = 
\begin{cases}
\frac{1}{\sum\limits_{r=0}^{\infty}\frac{|\alpha|^{2r}}{(3r)!}} \sum\limits_{n=0}^{\infty}
\frac{|\alpha|^{2n+2}}{(3n+2)!} & for \  $j=0$, \cr
\frac{1}{\sum\limits_{r=0}^{\infty}\frac{|\alpha|^{2r}}{(3r+1)!}} \sum\limits_{n=0}^{\infty}
\frac{|\alpha|^{2n}}{(3n)!} & for \ $j=1$, \cr
\frac{1}{\sum\limits_{r=0}^{\infty}\frac{|\alpha|^{2r}}{(3r+2)!}} \sum\limits_{n=0}^{\infty}
\frac{|\alpha|^{2n}}{(3n+1)!} & for \ $j=2$  .
\end{cases}
\end{eqnarray}
Plots of the uncertainty products $(\Delta x)_j(\Delta p)_j$ for $j=0,1,2$ are shown in Figure~\ref{dxdpj0}. 

It is important to explore the completeness relation in each subspace ${\cal H}_j$: 
\begin{eqnarray}
& \int \vert \alpha \rangle_j \, {}_j\langle \alpha \vert d \mu_j(\alpha) = I_j , 
\quad j=0,1,2, 
\label{completeHj}
\end{eqnarray}
where $I_j$ is the identity operator on ${\cal H}_j$ and
\begin{eqnarray}
& d \mu_j(\alpha) = \frac{1}{\pi\vert\alpha\vert} 
\left(\sum\limits_{r=0}^{\infty}\frac{|\alpha|^{2r}}{(3r+j)!}\right)
f_j(|\alpha|^2) d|\alpha| d \varphi .
\end{eqnarray}
If $f_j(x)$ 
satisfies
$\int_0^\infty x^{n-1} f_j(x) dx = \Gamma(3n + j + 1), \label{momentos}$
thus any state vector can be decomposed in terms of our CS.

Finally, the time evolution of a coherent state is quite simple,
$U(t)\vert \alpha \rangle_j = e^{-i(j+\frac12)t} \vert \alpha(t) \rangle_j , \ \alpha(t) 
= \alpha \, e^{-3it}$.

\begin{figure}
\begin{center}
\includegraphics[scale=0.66]{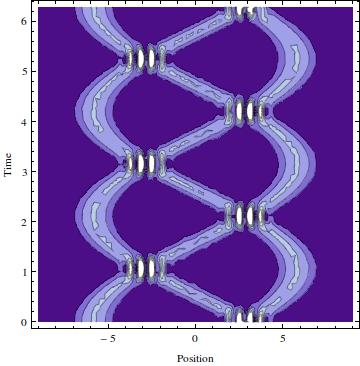}
\includegraphics[scale=0.66]{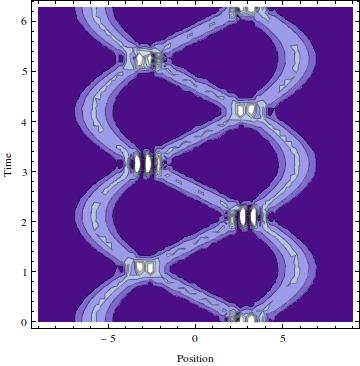}
\end{center}
\begin{center}
\includegraphics[scale=0.66]{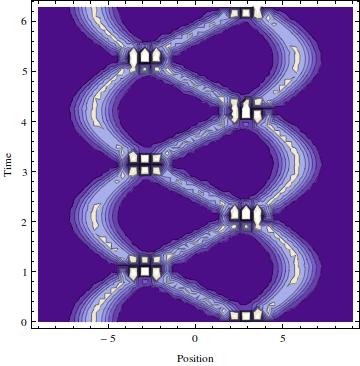}
\end{center}
\caption{Probability densities for the good, the bad and the ugly CS (left, right, down respectively).}
\label{pdj0}
\end{figure}

Let us consider next the non-normalized coherent states:
\begin{eqnarray}
&& \vert z\rangle = \sum\limits_{n=0}^\infty \frac{z^n}{\sqrt{n!}} \vert n\rangle, \label{estandarnon} 
\quad \vert z\rangle_j = \sum\limits_{n=0}^\infty \frac{z^{3n+j}}{\sqrt{(3n+j)!}} \vert 3n+j\rangle, 
\quad \alpha = z^3.  \label{multifotonnon}
\end{eqnarray}
The first state in Eq.~(\ref{estandarnon}) is a standard CS while the second ones are
the good, the bad and the ugly CS, also named three-photon CS \cite{D02}. Equation~(\ref{completeHj}) 
ensures that $\vert ze^{i2\pi j/3}\rangle$ can be written 
in terms of $\vert z\rangle_j, \ j=0,1,2$ \cite{Ce15}. 
Reciprocally, we can express $\vert z\rangle_j$  
in terms of $\vert ze^{i2\pi j/3}\rangle, \ j=0,1,2$:
\vskip-0.5cm
\begin{eqnarray}
&& \vert z\rangle_0 = N_0\left(\vert z\rangle + \vert e^{i2\pi/3}z\rangle + \vert e^{i4\pi/3}z\rangle\right), 
\label{multi-estandar0} \\
&& \vert z\rangle_1 = N_1\left(\vert z\rangle - e^{i\pi/3}\vert e^{i2\pi/3}z\rangle + e^{i2\pi/3}\vert e^{i4\pi/3}z\rangle\right), 
\label{multi-estandar1} \\
&& \vert z\rangle_2 = N_2\left(\vert z\rangle + e^{i2\pi/3}\vert e^{i2\pi/3}z\rangle + e^{i4\pi/3}\vert e^{i4\pi/3}z\rangle\right),
\label{multi-estandar2}
\end{eqnarray}
\vskip-0.2cm
\noindent i.e., the good, the bad and the ugly CS are superpositions of standard CS with complex labels 
$ze^{i2\pi j/3}$
defining an equilateral triangle on the complex plane. 
Expressions (\ref{multi-estandar0}-\ref{multi-estandar2}) are used to build the wave packets 
associated to $\vert z(t)\rangle_j, \ j=0,1,2$, 
whose probability densities as functions of $x$ and $t$ are shown in Figure~\ref{pdj0} 
\cite{Ce15}.

As we can see, the probability densities are periodic in time, with a period ($2\pi/3$) equal to one 
third of the period for a classical motion for the oscillator. This implies that the good, 
the bad and the ugly CS cannot describe semi-classical situations, i.e., they are intrinsically 
quantum states. It is worth to notice the existence of some other states which are strongly quantum, 
e.g., the even and odd CS \cite{RR00,D02,CL12,CF16}. 

\section{Conclusions}
\label{sec:4}

We have explored a realization of the second degree PHA in which the generators are the harmonic oscillator 
Hamiltonian and the ladder operators $a_g=a^3, \, a_g^+=(a^+)^3$. The three associated extremal states 
become physical eigenstates of $H$, and the ladders generated from them are of infinite length. In addition, 
these extremal states supply some solutions to the PIV equation. The search of the eigenstates of $a_g$ 
leads to three different sets, which here have been called the good, the bad and the ugly CS. Their period 
turns out to be a fraction ($1/3$) of the original period ($2\pi$) for the oscillator, indicating the 
strong quantum nature of such states. They could be important to describe the kind of interaction matter-radiation 
appearing in the so-called multiphoton quantum optics \cite{DDI06}.

\newpage

\bibliographystyle{unsrt}

\end{document}